\documentclass[aps,prb,twocolumn,superscriptaddress]{revtex4-2}
\usepackage[english]{babel}
\usepackage{bm}
\usepackage{amsmath}
\usepackage{graphicx}
\usepackage[colorlinks=true, allcolors=magenta]{hyperref}
\usepackage{diagbox}
\usepackage{multirow}
\usepackage{threeparttable}
\usepackage[table]{xcolor}

\begin{document}
	\title{Self-organised quantum dots in marginally twisted MoSe$_2$/WSe$_2$ and MoS$_2$/WS$_2$ bilayers}
	\author{V.~V.~Enaldiev}
	\affiliation{University of Manchester, School of Physics and Astronomy, Manchester M13 9PL, United Kingdom}
	\affiliation{National Graphene Institute, University of Manchester, Manchester M13 9PL, United Kingdom}
	
	\author{F. Ferreira}
	\affiliation{University of Manchester, School of Physics and Astronomy, Manchester M13 9PL, United Kingdom}
	\affiliation{National Graphene Institute, University of Manchester, Manchester M13 9PL, United Kingdom}
	
	\author{J. G. McHugh}
	\affiliation{University of Manchester, School of Physics and Astronomy, Manchester M13 9PL, United Kingdom}
	\affiliation{National Graphene Institute, University of Manchester, Manchester M13 9PL, United Kingdom}
	\author{V.I. Fal'ko}
	\affiliation{University of Manchester, School of Physics and Astronomy, Manchester M13 9PL, United Kingdom}
	\affiliation{National Graphene Institute, University of Manchester, Manchester M13 9PL, United Kingdom}
	\affiliation{Henry Royce Institute, University of Manchester, Manchester M13 9PL, United Kingdom}

	\begin{abstract}
		
	\end{abstract}
	\maketitle
	
	\textbf{Moir\'e superlattices in twistronic heterostructures are a powerful tool for materials engineering. In marginally twisted (small misalignment angle, $\theta$) bilayers of nearly lattice-matched two-dimensional (2D) crystals moir\'e patterns take the form of domains of commensurate stacking, separated by a network of domain walls (NoDW) with strain hot spots at the NoDW nodes. Here, we show that, for type-II transition metal dichalcogenide bilayers MoX$_2$/WX$_2$ (X=S, Se), the hydrostatic strain component in these hot spots creates quantum dots for electrons and holes. We investigate the electron/hole states bound by such objects, discussing their manifestations via the intralayer intraband infrared transitions.  The electron/hole confinement, which is strongest for $\theta<0.5^{\circ}$, leads to a red-shift of their recombination line producing single photon emitters (SPE) broadly tuneable around 1\,eV by misalignment angle. These self-organised dots can form in bilayers with both aligned and inverted MoX$_2$ and WX$_2$ unit cells, emitting photons with different polarizations. We also find that the hot spots of strain reduce the intralayer MoX$_2$ A-exciton energy, enabling selective population of the quantum dot states.}
	
 \section*{Introduction}
	The formation of minibands is a common moir\'e superlattice (mSL) effect \cite{cao2018correlated,cao2018unconventional,Yankowitz2019,lu2019,Xu2021,sharpe2019emergent,polshyn2020,Chen2019,chen2021,shen2020,park2021,cao2021nematicity,liu2020,Gadelha2021}, often related to a rigid rotation of one 2D crystal against the other. In general, the approximation of a rigid interlayer twist is valid for lattice-mismatched crystals or larger twist angles, where a short mSL period prohibits the formation of energetically preferential stacking domains of the two crystals. In contrast, for marginally (small-angle) twisted bilayers of crystals with very close lattice constants, the long period of the mSL offers sufficient space for creating preferential stacking areas. That is, the energy gain due to better adhesion can surmount the cost of intralayer strain in each of the constituent crystals. The reconstruction of small-angle twisted bilayers into a network of domains \cite{Enaldiev_PRL} [2H for antiparallel (AP) and 3R for parallel-oriented (P) bilayers] has been observed in various bilayers of transition metal dichalcogenides (TMDs) \cite{Weston2020,rosenberger2020,Sung2020,McGilly2020,Shabani2021}. The observed \cite{Weston2020,rosenberger2020,Sung2020,McGilly2020,Shabani2021} and theoretically modelled \cite{NaikPRL2018,CarrPRB2018,Enaldiev_PRL} structures feature hexagonal (for AP) and triangular (for P) NoDW with nodes hosting few nanometer areas of ''chalcogen-on-chalcogen'' stacking (X$_t$X$_b$), which are hot spots of the intralayer strain.    
	
	\begin{figure}[t]
		\includegraphics[width=1.0\columnwidth]{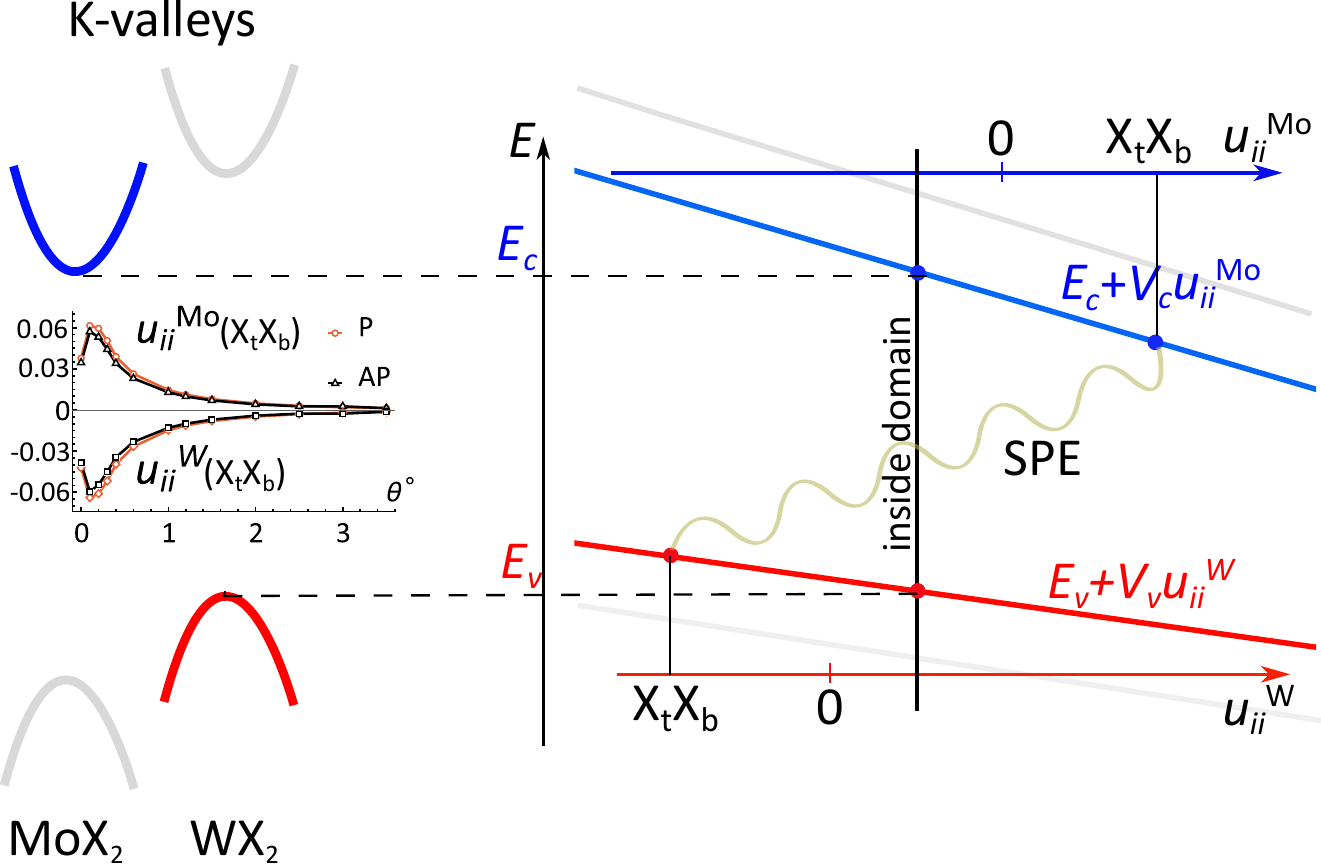}
		\caption{Variation of K-valley conduction/valence, $E_{c/v}$, band edges in MoX$_2$ (blue) and WX$_2$ (red) monolayers under hydrostatic strain with alignment of the bands corresponding to lattice-matched WX$_2$/MoX$_2$ heterobilayers. The vertical black solid line shows the value of hydrostatic strain of  individual layers inside domains. These values are opposite in sign with respect to those in X$_t$X$_b$ nodes, leading to formation of localised electron and hole states in self-organised QDs, which are suitable for SPE (wavy line). Inset shows twist angle dependence of hydrostatic components of strain in MoX$_2$ and WX$_2$ layers in X$_t$X$_b$ nodes of NoDW.}
		\label{fig:strain}
	\end{figure}
	
	Below we study the effects produced by these hot spots of strain in marginally twisted same-chalcogen heterobilayers MoX$_2$/WX$_2$. While domains/NoDW form in both homo- and heterobilayers, the in-plane intralayer deformations, $\bm{u}(\bm{r})$, in those two systems are qualitatively different. In homobilayers the formation of preferential stacking domains is brought about by twisting locally the crystals toward each other. As a result, the deformations at the domain walls are predominantly shear in character, that is, with ${\rm div}\, \bm{u}\equiv u_{ii} \to 0$, where $u_{ij}\equiv \frac{1}{2}(\partial_iu_j+\partial_ju_i)$ is a 2D strain tensor, and $u_{ii}$ is its trace. In perfectly aligned ($\theta=0^{\circ}$) heterobilayers, lattice mismatch ($\delta\approx 0.2\%$ for MoS$_2$/WS$_2$ and $\delta\approx 0.4\%$ for MoSe$_2$/WSe$_2$) requires an adjustment of the MoX$_2$ ($u_{ii}^{\rm Mo}\approx-\delta$ compression) and WX$_2$ ($u_{ii}^{\rm W}\approx\delta$ expansion) lattices towards each other inside the large area domains. This inflicts a few percent of hydrostatic compression of WX$_2$ and expansion of MoX$_2$ in X$_t$X$_b$ areas (NoDW nodes), quantified in Figs. \ref{fig:strain} and \ref{fig:band_edges_moire}, which, as we demonstrate below, create deep confinement potentials for charge carriers and interlayer excitons (iXs). Theoretical modelling \cite{Yu2017,Wu2018,Lu2019PRB,Brem2020,Guo2020} of localized iXs in WX$_2$/MoX$_2$ bilayer has been attempted earlier, however without taking into account strong lattice relaxation effects (which is applicable to structures with larger misalignment angles, $\theta>2^{\circ}$). This led to the underestimation of the depth of the size of the band edge variation for electrons and holes, as compared to what we find in this Letter, and with different positioning of band edge extrema across moir\'e supercell. In this work we focus on small-angle twisted WX$_2$/MoX$_2$ bilayers where lattice relaxation plays the critical role on trapping charge carriers, in particular due to a substantial hydrostatic strain component at NoDW nodes. Up to now, no optical studies have been reported on such small-angle ($\theta\leq 1^{\circ}$) bilayers, despite that the structural features of NoDW have been demonstrated using transmission electron microscopy \cite{Weston2020,rosenberger2020} and several spectroscopic studies \cite{Tran2019,Seyler2019,Baek2020,BrotonsGisbert2020} have been performed on bilayers with larger misalignment angles, where lattice reconstruction does not play such a dominant role as discussed below. 
	
	\section*{Results and Discussion}
	{\bf Modulation of band edges by strain and charge transfer}. Here, we single out the hydrostatic strain component because of the critical role it plays in determining the K-valley energies in MoX$_2$/WX$_2$ crystals. Several earlier experimental and density functional theory (DFT) studies \cite{Conley2013, Dhakal2017, Zollner2019} have agreed that conduction and valence band edges in TMD monolayers are strongly shifted by hydrostatic strain, but without much sensitivity to shear deformations. This trend is illustrated in Fig. \ref{fig:strain}. The corresponding shifts of conduction/valence band edges in MoX$_2$/WX$_2$ determine the energy of the interlayer exciton (iX). Inside domains formed by lattice reconstruction, hydrostatic strains compensate lattice mismatch between the layers: this slightly increases the band gap as compared to rigidly twisted bilayers without strain (Fig. \ref{fig:strain}). Compensating small deformations inside large domain areas, $u_{ii}^{\rm Mo,W}$ in X$_t$X$_b$ nodes have the opposite signs and much larger magnitudes as compare to $u_{ii}^{\rm Mo,W}$ inside domains. This strongly decreases layer-indirect band gap and determines deep confinement potentials for both electrons and holes, leading to the appearance of SPEs.  
	
	\begin{figure}
		\includegraphics[width=1.0\columnwidth]{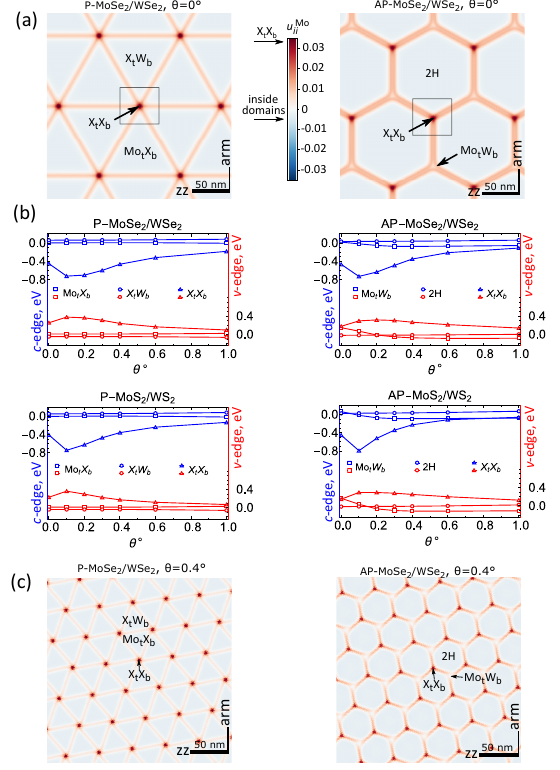}
		\caption{Lattice reconstruction and local conduction and valence band edge modulations at K-valleys in twisted MoX$_2$/WX$_2$ bilayers. Panels (a) and (c) show distribution of the hydrostatic strain, $u_{ii}^{\rm Mo}$, in MoSe$_2$ layer of P- and AP-MoSe$_2$/WSe$_2$ bilayers with $\theta=0^{\circ}$ and $\theta=0.4^{\circ}$, respectively. Squares in (a) show areas used to plot band edge profile in Fig. \ref{Fig:XX_band_edge}. Arrows on the scalebar indicate the  values of $u_{ii}^{\rm Mo}$ inside domains and at NoDW nodes. (b) Modulations of conduction ($c$) and valence ($v$) band edges as a function of twist angle for high symmetry stackings of moir\'e supercell, indicated by different symbols, for P/AP- MoSe$_2$/WSe$_2$ and MoS$_2$/WS$_2$ bilayers. Matching MoX$_2$ and WX$_2$ lattices inside the large area domains determines homogeneous strain which is accounted for through offsetting of MoX$_2$ and WX$_2$ strain axes in Fig. \ref{fig:strain}.}
		\label{fig:band_edges_moire}
	\end{figure}
	
	
	The intralayer strain ($u_{ii}^{\rm Mo/W}$) maps in Fig. \ref{fig:band_edges_moire} were computed using a multiscale modelling approach \cite{Enaldiev_PRL}, tested in the detailed comparison with the STEM microscopy data \cite{Weston2020}. This approach starts with the computation of stacking-dependent MoX$_2$/WX$_2$ adhesion energy, $\mathcal{W}$, followed by the parametrization of interpolation formulae \cite{Enaldiev_PRL} for its dependence on the interlayer lateral offset, $\bm{r}_0$. For both P and AP bilayers, energetically favourable stackings are those with the largest lateral separation between chalcogens. These stackings are Mo$_t$X$_b$ and X$_t$W$_b$ for P-orientation and 2H for AP-orientation. Note that X$_t$X$_b$ stacking is unfavourable energetically, and its interlayer distance swells \cite{Enaldiev_PRL} by up to $\approx$0.5\,\AA. By combining interpolation formulae for $\mathcal{W}(\bm{r}_0)$ \cite{Enaldiev_PRL} where we use local lateral offset, 
	\begin{equation}\label{Eq:}
		\bm{r}_0(\bm{r})=\delta\cdot\bm{r}+\theta\hat{z}\times\bm{r}+\bm{u}^{\rm Mo} - \bm{u}^{\rm W},    
	\end{equation}
	with elasticity theory and minimizing total energy of the bilayer across its mSL (which period, $\ell \approx a/\sqrt{\theta^2+\delta^2}$, is fixed by the twist angle and lattice mismatch between the crystals), we compute the deformation fields. 
	
	In Fig. \ref{fig:band_edges_moire}(a,c), we present $u^{\rm Mo}_{ii}$ maps for MoSe$_2$/WSe$_2$ bilayers with $\theta=0^{\circ}$ and $\theta=0.4^{\circ}$, where the preferential stacking domains (triangular/hexagonal for P/AP-bilayer) are separated by networks of dislocation-like domain walls with X$_t$X$_b$ stacking at the NoDW nodes \cite{Enaldiev_PRL}. Note that for small twist angles the hydrostatic component of strain persists, now combined with shear deformations (similar to those in twisted homobilayers).
	
	To incorporate strain into the shifts of conduction/valence band edges, $\delta \varepsilon_{c/v}$, we performed DFT modelling of the TMD band structures using Quantum ESPRESSO \cite{QE-2009}. For each material, we considered biaxial strain in the range of $\pm 2$\%, fully relaxing atomic positions in the monolayer with Vanderbilt PBE GBRV ultrasoft pseudopotentials \cite{Garrity2014}, a wavefunction cut-off of $E_{cut}$ = 50 Ry, and a $20 \times 20 \times 1$ \textit{k}-point grid, sampled according to the Monkhorst-Pack algorithm \cite{Monkhorst1976}. Spin-orbit coupling was included by a norm-conserving fully-relativistic pslibrary PBE PAW pseudopotentials \footnote{We used the relevant relativistic pseudopotentials from http://www.quantum-espresso.org.} with $E_{cut}$ = 80 Ry.

\begin{figure*}
		\includegraphics[width=2.0\columnwidth]{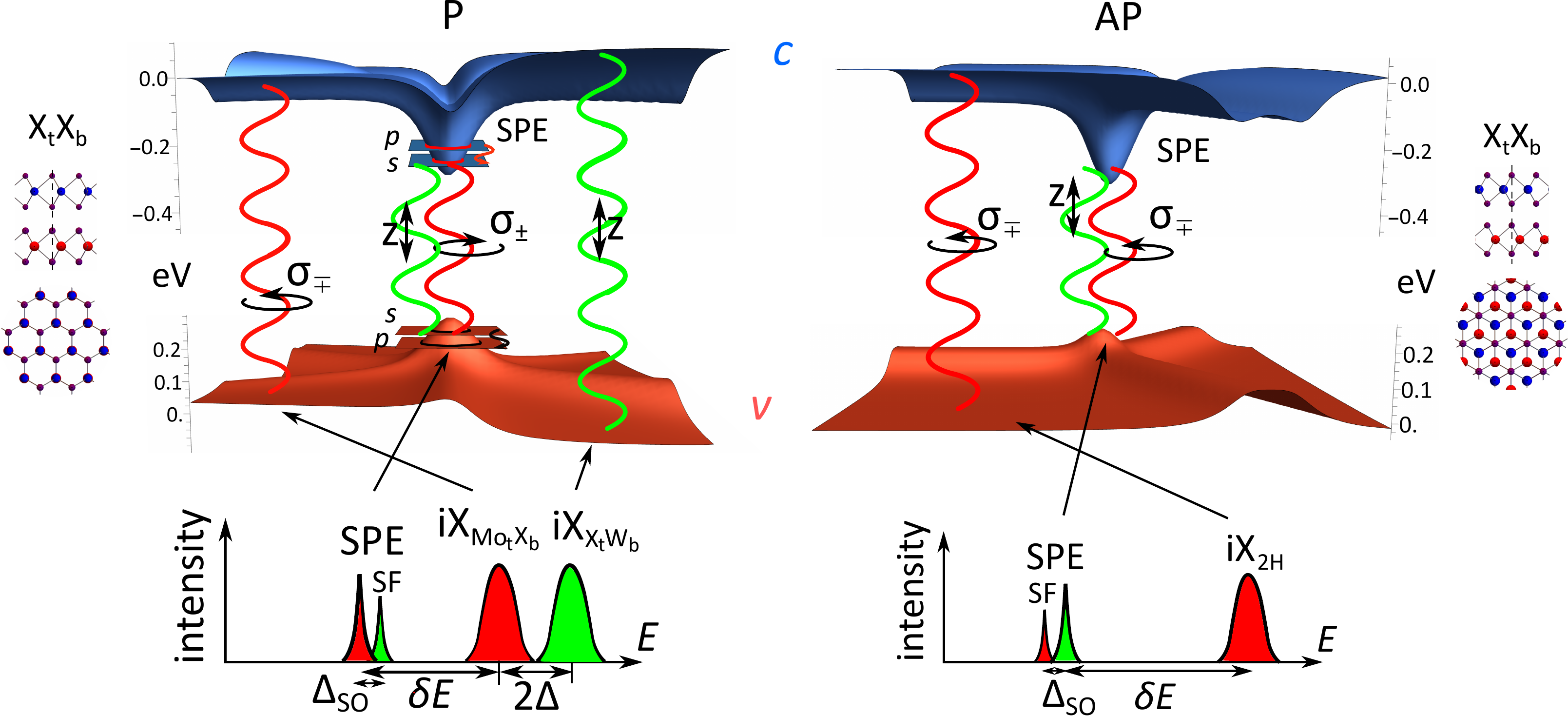}
		\caption{Self-organised quantum dots and spectral features of SPE and iX. (Top) Conduction ($c$) and valence ($v$) band edge profiles in vicinity of X$_t$X$_b$ nodes of NoDW in reconstructed P-  and AP-MoX$_2$/WX$_2$ bilayers with $\theta=0^{\circ}$. Colors of wavy lines encode polarisations of emitted light in $\pm$K-valleys: red for circular and green for $z$-polarisation. Upper/lower subscript of circular polarisation ($\sigma_{\pm}$ or $\sigma_{\mp}$) indicates helicity of light emitted in $+$K/$-$K-valleys. Left and right bottom panels show sketches of predicted optical spectra in marginally twisted P- and AP-MoX$_2$/WX$_2$ bilayers, respectively.} 
		\label{Fig:XX_band_edge}
	\end{figure*}

	The computed variations of all band edges can be described as $\approx V_{v/c} u_{ii}$, with $V_{c}^{\rm MoS_2}=-12.45$\,eV, $V_{v}^{\rm WS_2}=-5.94$\,eV, $V_{c}^{\rm MoSe_2}=-11.57$\,eV, $V_{v}^{\rm WSe_2}=-5.76$\,eV, which are quoted for the relevant bilayer bands. Using these values, we compute, 
	\begin{equation}\label{Eq:band_edges}
		\delta\varepsilon_{v/c}(\bm{r})=V_{v/c}u_{ii}^{\rm W/Mo}(\bm{r})-e\phi^{\rm W/Mo}_{\rm piezo}(\bm{r})  \pm \frac 12\Delta(\bm{r}),
	\end{equation}
	taking into account strain-dependent piezoelectric potential \cite{Enaldiev_PRL}, $-e\phi^{\rm W/Mo}_{\rm piezo}(\bm{r})$, and offset-dependent potential drop, $\Delta(\bm{r})$, due to interlayer charge transfer (for details see section S1 in Supplementary Information (SI)). The first term in Eq. \eqref{Eq:band_edges} represents the effect of hydrostatic component of the intralayer strain which was missed in the previous analysis of the same systems \cite{Enaldiev_2021}. The twist-angle dependences of the computed band edge energies $\delta \varepsilon_{c/v}$ for three selected stacking areas (Mo$_t$X$_b$, X$_t$W$_b$, X$_t$X$_b$ for P and 2H, Mo$_t$W$_b$, X$_t$X$_b$ for AP) are plotted in Fig. \ref{fig:band_edges_moire}(b). These figures suggest that X$_t$X$_b$ regions are potential wells for electrons and holes and these wells are the deepest for $\theta\approx\delta$. Based on that we describe the NoDW nodes as trigonally warped quantum dots (QDs), with band edge profiles exemplified in Fig. \ref{Fig:XX_band_edge}. These QDs are sufficiently deep to accommodate at least two size-quantized states for electrons/holes which retain their distinct $s$ ($L_z=0$) and $p$ ($L_z=\pm1$) characteristics due to the $\hat{C}_3$-symmetry of the dots.
	
	Note that QD formation in marginally twisted structures qualitatively differs from band energy profiles in stronger misaligned P-bilayers with $\theta\geq 2^{\circ}$, where the band edges at K-valley shift into Mo$_t$X$_b$ stacking areas, see Extended data Fig. \ref{fig5:eg_modulation}. This crossover agrees with the findings of Refs.  \cite{Yu2017,FuPRB2018,MalicNanoLett2020}. In addition, for MoS$_2$/WS$_2$ with $\theta\approx1.8^{\circ}$ and MoSe$_2$/WSe$_2$ with $\theta\approx2.4^{\circ}$ the energy profile for interlayer interband exciton resembles an antidot superlattice more than an array of QDs. This contrasts with the persistence up to $\theta\sim 3.5^{\circ}$ of shallow QD arrays for both electrons and holes based at X$_t$X$_b$ areas in AP-bilayers.
	
\begin{figure}
		\includegraphics[width=1.0\columnwidth]{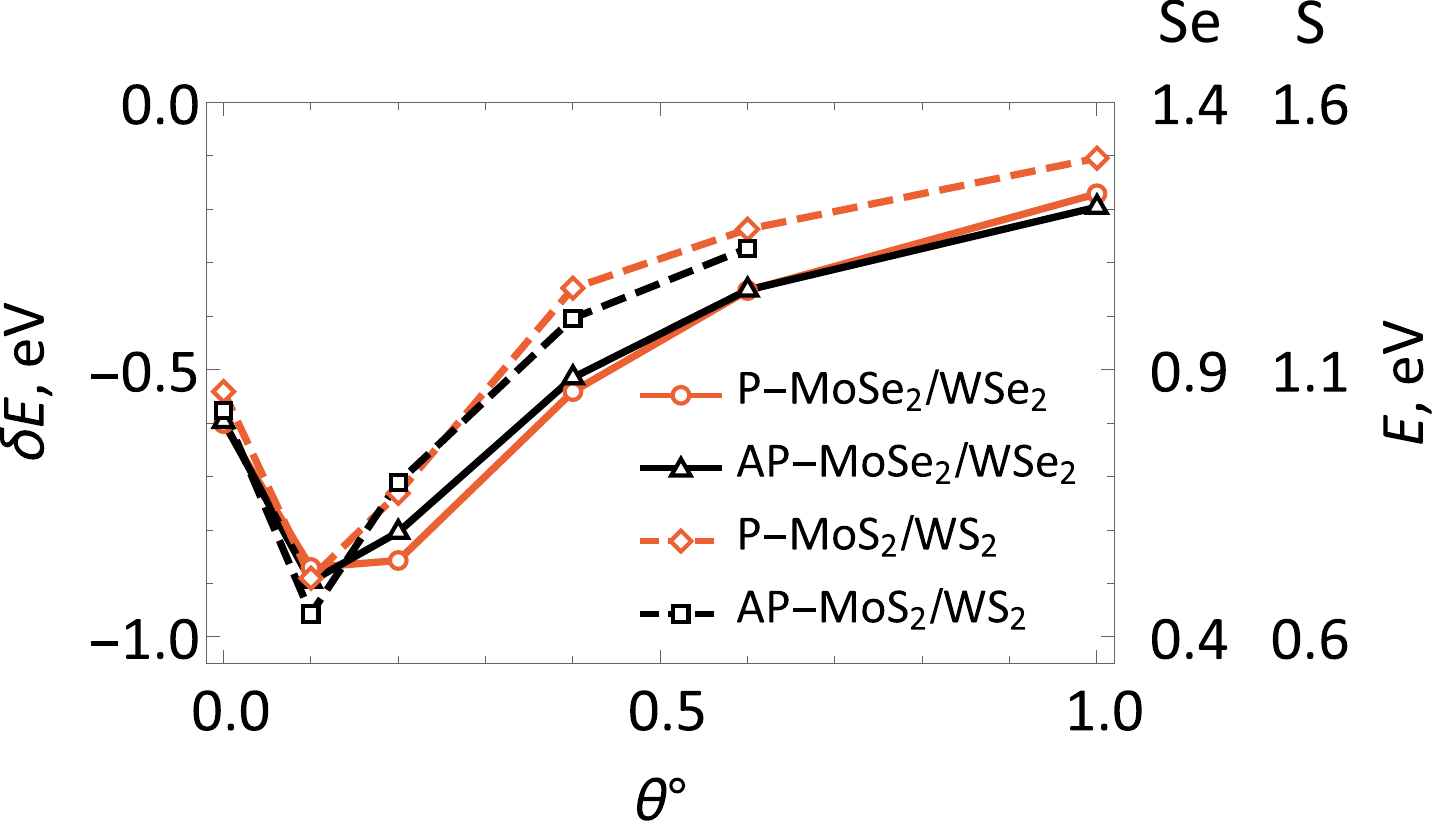}
		\caption{Energies of SPEs in P/AP-WX$_2$/MoX$_2$ bilayers shown as a shift from the lower iX energy inside the domains (left-side axis common for all bilayers) and as an absolute value (right-side axis specified separately for selenides and sulfides). The latter  were determined as $E=\delta E+1.4$\,eV and $E=\delta E+1.6$\,eV for selenide (Se) and sulfide (S), respectively. These estimates are based on calculated binding energies of iXs $\approx 65$\,meV (see details in SI) and measured positions of photoluminescence peaks of iXs in MoSe$_2$/WSe$_2$ \cite{Baek2020} and MoS$_2$/WS$_2$ \cite{Kiemle2020}.}
		\label{Fig:SPEenergies}
	\end{figure}

	{\bf Spectral characteristics of self-organised QDs.} Spectral features of the interlayer interband emissions of self-organised QDs of marginally twisted bilayers are sketched on the bottom insets in Fig. \ref{Fig:XX_band_edge}. 
	\,\,Energy separation, $\delta E$, between the QD transition and the iX inside the domains was computed as,
	\begin{align}\label{EQ:deltaE}
		\delta E =&\,  \varepsilon_e^{(s)} - \varepsilon_h^{(s)} -  E_{\rm iX} \\ 
		&- \int\int d^2\bm{r}d^2\bm{r}'\left|\psi_e^{(s)}(\bm{r})\right|^2\left|\psi_h^{(s)}(\bm{r}')\right|^2V_{eh}(\bm{r}-\bm{r}'). \nonumber
	\end{align}
	Here, $E_{\rm iX}$ is the iX energy and $\varepsilon_{e/h}^{(s)}$ are the energies of the electron/hole $s$-states $\psi_{e/h}^{(s)}$ inside quantum well. We also take into account the interlayer e-h attraction, $V_{eh}$, screened by the in-plane susceptibility of TMDs and hBN environment  \cite{DanovichPRB} (see details in Sections S2, S3 in SI). The computed dependences of $\delta E(\theta)$ for MoX$_2$/WX$_2$ bilayers (X=Se,S) are shown in Fig. \ref{Fig:SPEenergies}. We find that the QD line can be tuned across a 0.8-1.2\,eV spectral interval (telecom range) for $0.3^{\circ}\leq\theta\leq 1^{\circ}$ in MoSe$_2$/WSe$_2$ and for $0^{\circ}\leq\theta\leq 0.5^{\circ}$ in MoS$_2$/WS$_2$. In Fig. \ref{Fig:SPEenergies} the computed data for electrons in AP-MoS$_2$/WS$_2$ are terminated at $\theta=0.6^{\circ}$, because for a larger misalignment the K-valley conduction band profile starts resembling an antidot lattice with maxima at the 2H domains (see Fig. \ref{fig:band_edges_moire}(b)).
	
	 {\bf Polarization and spin selection rules.} Additional information, displayed in Fig. \ref{Fig:XX_band_edge} and gathered in Table \ref{Tab:polar}, concerns the polarizations of SPEs and iXs inside domains and fine structure related to a $\Delta_{\rm SO}$-splitting \cite{Mak2012,Echeverry2016,Wang2017,Liu2013,Kormnyos2015} between the spin-flip (SF) and spin-conserving interband transitions inside the QDs. We note that, in each of the two $\pm$K-valleys, the hole spin at the band edge is determined by the spin-valley locking in WX$_2$, whereas conduction band in MoX$_2$ is characterized by spin-orbit splitting $\Delta_{\rm SO}$. Also, the iX emission from the inner part of domains, shown in Fig. \ref{Fig:XX_band_edge}, differs for P and AP bilayers. For AP-bilayers we expect a single line of circularly polarised iX emission. For P-bilayers iX energies and polarisations are different for Mo$_t$X$_b$ and X$_t$W$_b$ domains, with the energy splitting determined by the interlayer charge transfer \cite{Ferreira2021,Enaldiev_2021,Weston2022} and circular (in Mo$_t$X$_b$) {\it vs} linear (in X$_t$W$_b$) polarization, established in Ref. \cite{Yu2018_2DMat}. In Table \ref{Tab:polar} we highlight the most prominent SPE transition which happens to be related to the spin-conserving recombination in self-organised QDs in P-bilayers with approximately one hundred times weaker intensity than that of intralayer A-exciton (AX) in MoX$_2$ layer determined by the ratio of corresponding interband matrix elements. To mention, DFT modelling suggests that lattice matching inside domains promotes direct-to-indirect band gap crossover for electrons towards Q-valley and most importantly for holes towards $\Gamma$-valley \cite{Kiemle2020} (see also Section S6 in SI).

	\begin{table*}
		\caption{Polarization of emitted photons by SPE, iX and AX in MoX$_2$, and strength of the emission characterised by interband velocity matrix element, $|v^{cv}_{\pm,z}|$ (in \mbox{$10^6\times {\rm cm/s}$}, $v_{\pm}=v_x\pm iv_y$) computed with Quantum ESPRESSO (for details see Sections S4 and S5 in SI). $\sigma_+$ ($\sigma_{-}$) designates photons with clockwise (counter clockwise) polarization outgoing in positive direction of spin quantization axis. $z$ stands for linear out-of-plane polarization. Here, upper (lower) index corresponds to emission from $+$K ($-$K) valley.  For this analysis we employed angular momentum conservation for K-point Bloch states (we used center of coordinate set at a chalcogen atom).}
		{
			\begin{tabular}{c|c|ccc|cc}
				\hline
				\hline
				&	AX MoX$_2$ & \multicolumn{3}{c|}{P} & \multicolumn{2}{c}{AP} \\
				& & SPE$_{\rm X_tX_b}$ & iX$_{\rm Mo_tX_b}$  &  iX$_{\rm X_tW_b}$ & SPE$_{\rm X_tX_b}$ & iX$_{\rm 2H}$  \\
				\hline
				\multirow{2}{*}{ground ($E$)}&  & \multicolumn{3}{c|}{spin-conserving} & \multicolumn{2}{c}{spin-flip} \\
				& $\sigma_{\pm}$ & $\sigma_{\pm}$ &  $\sigma_{\mp}$ & $z$ & $\sigma_{\mp}$& $\sigma_{\pm}$\\
				S  & 113 & \cellcolor{yellow!50} 6.42 & 4.57 & 2.53 &  0.48 & 2.28 \\
				Se & 97 & \cellcolor{yellow!50} 7.03 & 5.30 & 3.73 & 0.54 & 3.49 \\
				\hline
				\multirow{2}{*}{excited ($E+\Delta_{\rm SO}$)} & & \multicolumn{3}{c|}{spin-flip} & \multicolumn{2}{c}{spin-conserving} \\
				& & $z$ & $\sigma_{\pm}$ & $\sigma_{\mp}$ & $z$ & $\sigma_{\mp}$ \\ 
				S & & 0.49 & 2.82 & 0.12 & 1.18 & 10.30 \\
				Se & & 0.74 & 5.01 & 0.10 & 1.89 & 10.64 \\
				\hline
				\hline
			\end{tabular}
			\label{Tab:polar}}
	\end{table*}

	\begin{figure}
		\includegraphics[width=1.0\columnwidth]{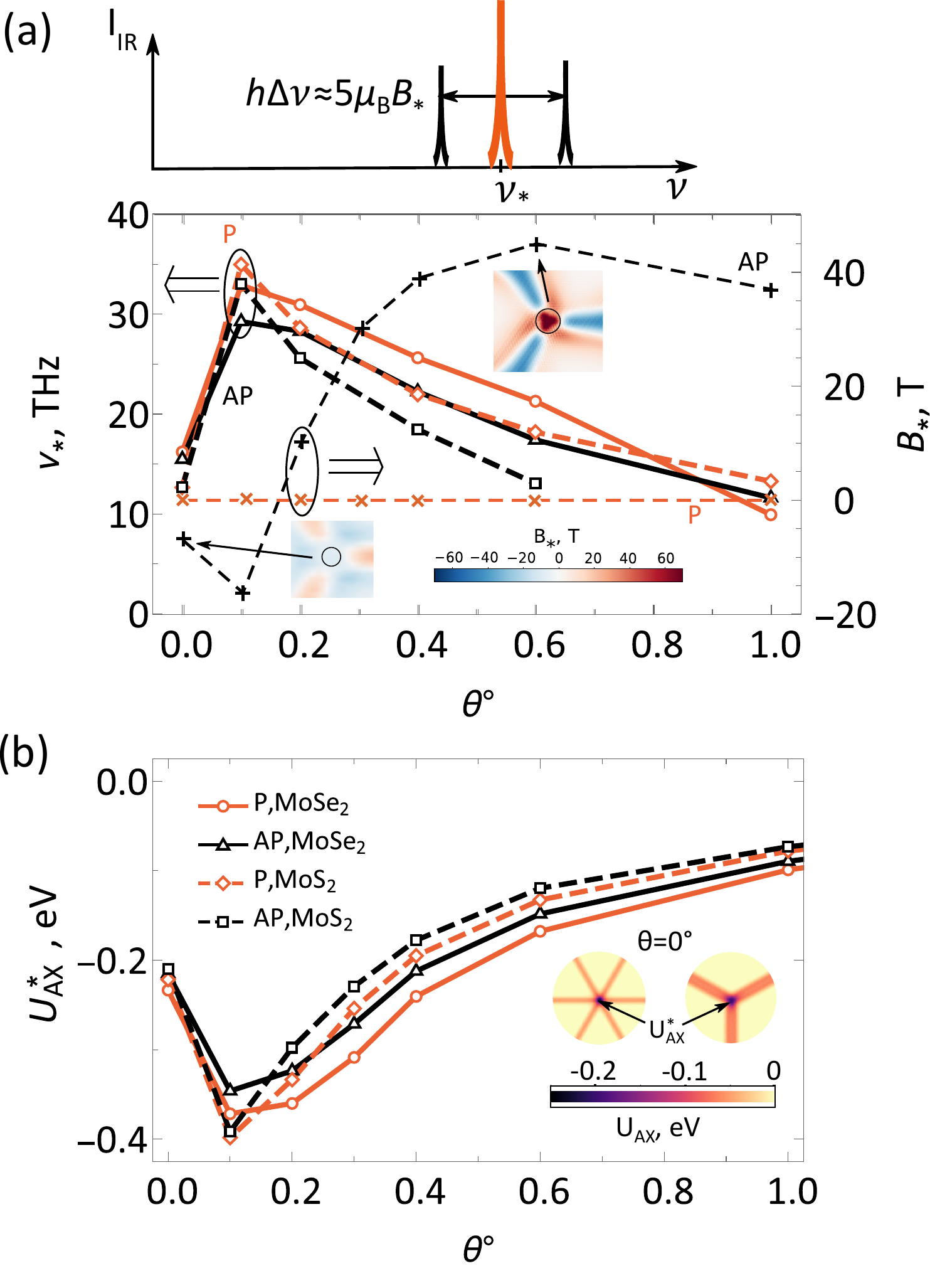}
		\caption{Intralayer infrared features of QDs and intralayer (MoX$_2$) band gap variation with marking of materials and orientations of the bilayers shown on panel (b). (a) Infrared line, $\nu_*$, of a QD in P- and AP-MoX$_2$/WX$_2$ bilayers, with a line splitting due to pseudomagnetic field, $B_*$, in AP-structures. Twist angle dependence of QD $s-p$-transition frequency was analysed based on Eq. \ref{Eq:band_edges} and $B_*$ was estimated for conduction band electrons in MoS$_2$ using parameters from Ref. \cite{Rostami2015}. Pseudomagnetic field maps around X$_t$X$_b$ nodes of NoDW for $\theta=0^{\circ}$ and $\theta=0.6^{\circ}$ are shown in the insets. Pseudomagnetic fields of similar magnitude are expected for holes in MoS$_2$/WS$_2$ bilayers and both electrons and holes for MoSe$_2$/WSe$_2$ bilayers. (b) Depth of a potential well, $U^*_{\rm AX}$, confining intralayer A-excitons in MoX$_2$, created by hydrostatic strain around X$_t$X$_b$ nodes of NoDW. Inset shows anisotropy of the well profile for P- and AP-MoSe$_2$/WSe$_2$ bilayers with $\theta=0^{\circ}$. }
		\label{fig5:AX_potential}
	\end{figure}
	
	{\bf Intralayer transitions in self-organised QDs.} In addition, intraband $s-p$ optical transitions for electrons/holes trapped in QDs give rise to infrared (IR) features with energies shown in Fig. \ref{fig5:AX_potential}(a) for twist angles $0^{\circ}\leq\theta\leq 1^{\circ}$. In AP-bilayers, X$_t$X$_b$ NoDW nodes also feature spikes of pseudomagnetic field $B_*$ \cite{Enaldiev_PRL} characteristic of multivalley semiconductors with a strongly inhomogeneous strain \cite{Iordanskii1985,Suzuura2002,Rostami2015}. These pseudomagnetic fields have opposite signs for electrons in $\pm$K-valleys, splitting (by $\hbar\Delta\nu=5\mu_BB_*$) the QD $s-p$ transitions into circularly polarized doublets, as sketched in Fig. \ref{fig5:AX_potential}(a). Such an infrared transition can be used to manipulate the state of the SPE, by exciting either electron or hole into their respective QD $p$-states.
	
	The intralayer band gap variation, due to the hydrostatic strain at X$_t$X$_b$ nodes reduces/increases the energy of the intralayer AX in MoX$_2$/WX$_2$. In Fig. \ref{fig5:AX_potential}(b) we show  that this results in a $\sim$100 meV potential well for AX in MoX$_2$ exactly over the self-organised QD position. The red-shift of the MoX$_2$ AX confined in such a well can be used for selective population of the QD states, upon the relaxation of photoexcited hole in MoX$_2$ layer into its bound state in the QD in WX$_2$. 
	
	\section*{conclusions}
	Overall, hot spots of hydrostatic strain at the nodes of domain wall network, generated by the lattice reconstruction in marginally twisted MoX$_2$/WX$_2$ bilayers, form a nanoscale array of quantum dots for electrons and holes, which may be operated as single-photon emitters. 
	Based on the presented analysis, we propose that the SPE spectrum can be tuned by the choice of the twist angle over a broad range (including telecom for MoSe$_2$/WSe$_2$ bilayers), and the electron/hole state in these QDs can be manipulated via intra-band $s-p$ transitions using THz radiation. 
	The data on the optical oscillator strength of the interlayer interband transitions in such QDs, Table \ref{Tab:polar}, suggest  that the brightest would be SPEs in marginally twisted biayers with parallel orientation of MoX$_2$ and WX$_2$ unit cells. Ratio of intra- and inter-layer interband velocity matrix elements also suggest that the recombination rate of QD-localised excitons is about $\sim 1\%$ of the recombination rate of the intralayer A-exciton in MoX$_2$;  as the latter was found in Refs. \cite{Poellmann2015,Dey2016,Jakubczyk2016,Cadiz2017} to be $\sim 1/300-1/200$\,fs$^{-1}$, this would set a 100 MHz possible repetition rate for SPEs in these self-organised QDs. Note that areal density of these SPEs is $\sim$10$^{11}$\,cm$^{-2}$, which is 100 times higher than the density of quantum emitters in patterned TMD monolayers \cite{PalaciosBerraquero2017,Luo2018} and that a red-shift of the A-exciton in MoX$_2$, due to the same hot spots of strain, would enable  selective population of the QD states for the optical pumping of the self-organised SPEs. 
 
	Finally, we note that in real samples there is usually inhomogeneity of domain structure caused by smooth strain introduced during sample transfer process. Such inhomogeneity spread a spectral range of SPEs available on a single wafer, which can be used to produce a wealth of SPE devices operating in complementary spectral intervals provided by different parts of a single large-area WX$_2$/MoX$_2$ bilayer. 
	
	{\bf Acknowledgements.} This work was supported by EC-FET European Graphene Flagship Core3 Project, EC-FET Quantum Flagship Project 2D-SIPC, EPSRC grants EP/S030719/1 and EP/V007033/1, and the Lloyd Register Foundation Nanotechnology Grant.
	
	\renewcommand{\figurename}{Extended data Fig.}
	\setcounter{figure}{0}
	\begin{figure}
		\includegraphics[width=1.0\columnwidth]{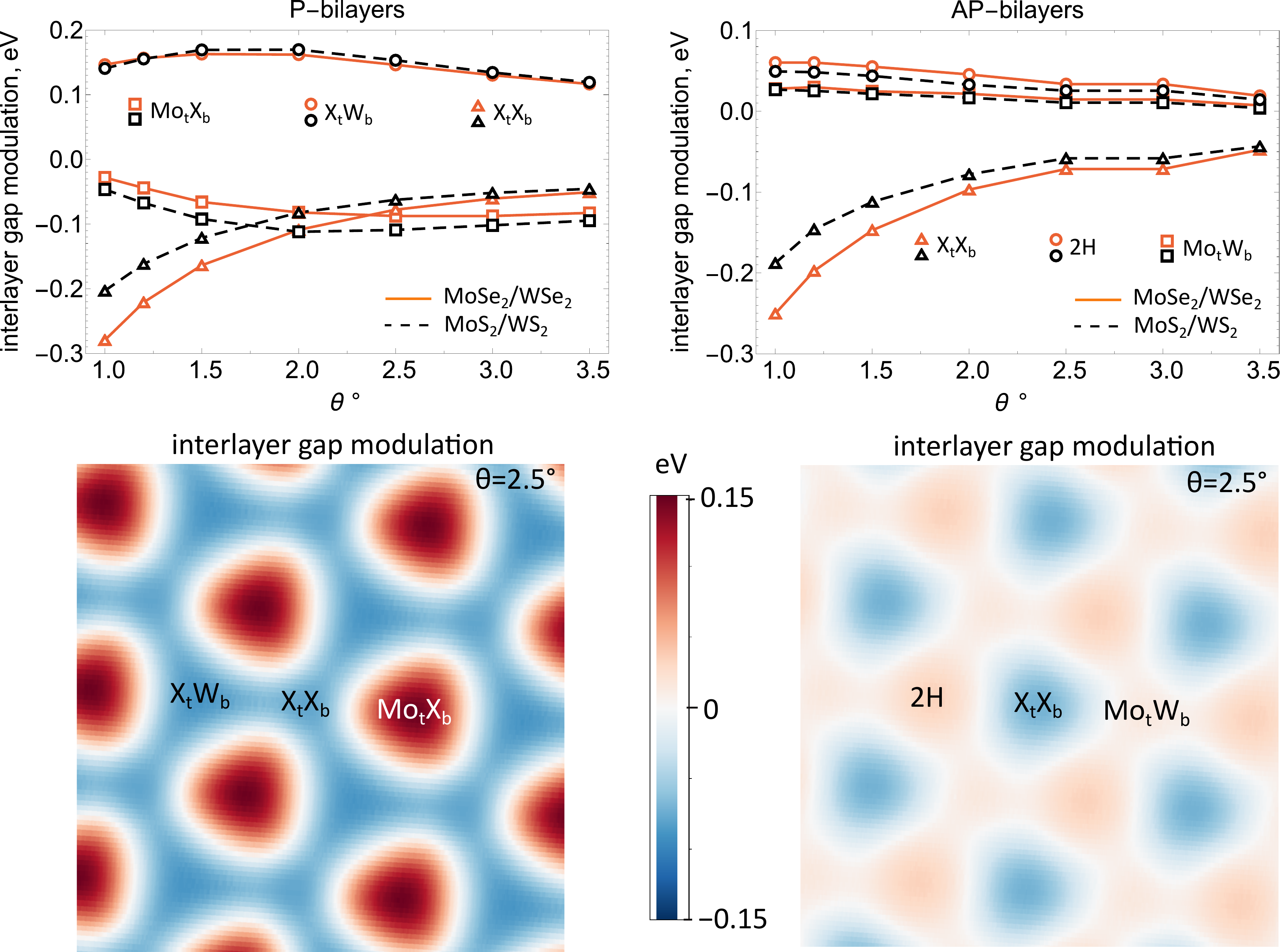}
		\caption{(Top) Twist-angle-dependences of interlayer band gap at K-valleys at high symmetry stacking areas of moir\'e supercells of P- and AP-MoX$_2$/WX$_2$ bilayers with $1^{\circ}\leq \theta\leq 3.5^{\circ}$. For P-bilayers with $\theta\gtrsim 2^{\circ}$, we find a crossover from an array of QDs located in X$_t$X$_b$ nodes to an antidot array with shallow minima at X$_t$W$_b$ areas and high peaks at Mo$_t$X$_b$ areas, while for AP-bilayers there is no such transition. (Bottom) Real space maps showing crossover (towards larger misalignment) of the interlayer band gap for MoS$_2$/WS$_2$ bilayers with $\theta= 2.5^{\circ}$: an antidot array for P-bilayers and shallow QD array for AP-bilayers.}
		\label{fig5:eg_modulation}
	\end{figure}

\bibliography{bibl}
	
\end{document}